\documentstyle[epsfig]{elsart}
\begin{document}
\date{16 April 1997}
\begin{frontmatter}
\title{Added noise in homodyne measurement of field-observables}
\author{G. Mauro D'Ariano and Matteo G. A. Paris}
\address{Dipartimento di Fisica 'Alessandro Volta' dell'Universit\'a 
         degli Studi di Pavia \\ Istituto Nazionale di Fisica della
         Materia -- Sezione di Pavia \\via A. Bassi 6, I-27100, Pavia,
         Italia.}
\begin{abstract}
         Homodyne tomography provides a way for measuring generic
         field-operators. Here we analyze the determination of the
         most relevant quantities: intensity, field, amplitude and
         phase. We show that tomographic measurements are affected
         by additional noise in comparison with the direct detection
         of each observable by itself. The case of of coherent states
         has been analyzed in details and earlier estimations of
         tomographic precision are critically discussed.
\end{abstract}
\end{frontmatter}
\section{Introduction}
One of the most exciting developments in the recent history of quantum
optics is represented by the so-called {\em Homodyne Tomography}, namely
the homodyne detection of a nearly single-mode radiation field while
scanning the phase of the local oscillator \cite{ray,tom,ulf,dar}. 
>From a tomographic data record, in fact, the density matrix elements can 
be recovered, thus leading to a complete characterization of the quantum 
state of the field. This is true also when not fully efficient photodetectors 
are involved in the measurement, provided that quantum efficiency is
larger than the threshold value $\eta=1/2$. \par
In homodyne tomography a general matrix element is obtained
as an expectation value over homodyne outcomes at different phases. In
formula
\begin{eqnarray}
\langle\psi |\hat{\varrho}|\varphi\rangle = \int^{\pi}_0 
\frac{d\phi}{\pi}
\int^{+\infty}_{-\infty}\!\!dx\:p_{\eta}(x;\phi)\:\langle\psi |
K_{\eta}(x- \hat{x}_{\phi})|\varphi\rangle\, ,\label{bas1}
\end{eqnarray}
where $p_{\eta}(x;\phi)$ is the probability density of the homodyne
outcome $x$ at phase $\phi$ for quantum efficiency $\eta$ and the integral
kernel is given by
\begin{eqnarray}
K_{\eta} (x) = \frac{1}{2} \hbox{Re} \int^{+\infty}_0 \!\!dk\,
k\,\exp \left( \frac{1-\eta}{8\eta}k^2+ikx\right) \, .\label{bas2}
\end{eqnarray}
While the kernel in Eq. (\ref{bas2}) is not even a tempered
distribution, its matrix elements can be bounded functions depending
on the value of $\eta$. This is the case of the number representation
of the density matrix, for which the "pattern function"
\begin{eqnarray}
f_{n,n+d}^{(\eta )}(x,\phi) \equiv \langle n|K_{\eta} (x-\hat{x}_{\phi})|
n+d\rangle  
\label{bas3}\;.
\end{eqnarray}
can be expressed as a finite linear combination of parabolic cylinder
functions \cite{ulf}.
\par\noindent 
As it comes from the experimental average
\begin{eqnarray}
\overline{\varrho_{n,m}}= \int^{\pi}_0 \frac{d\phi}{\pi}
\int^{+\infty}_{-\infty} \!\! dx\, p_{\eta}(x;\phi)\: 
f_{n,n+d}^{(\eta )}(x,\phi)       
\, ,\label{samp1}
\end{eqnarray}
the tomographic determination $\overline{\varrho_{n,m}}$ for the matrix
element  $\varrho_{n,m}\equiv\langle n| \hat \varrho | m \rangle$ is
meaningful only when its confidence interval is specified. This is
defined, according to the central limit theorem, as the rms value
rescaled by the number $N$ of data. As $\varrho_{n,m}$ is a complex
number, we need to specify two errors, one for the real part and one for
the imaginary part respectively. For the real part one has
\begin{eqnarray}
\hbox{Re}\:\varepsilon_{n,m}=\frac{\overline{\Delta\hbox{Re}
\varrho_{n,m}}}{N} =\frac{\left\{\overline{\hbox{Re}\varrho^2_{n,m}} -
\left[\hbox{Re}\overline{\varrho_{n,m}}\right]^2 \right\}^{1/2}}{N}
\, ,\label{samp2}
\end{eqnarray}
where
\begin{eqnarray}
\overline{\hbox{Re}\varrho^2_{n,m}} = \int^{\pi}_0 \frac{d\phi}{\pi}
\int^{+\infty}_{-\infty} \!\! dx\, p_{\eta}(x;\phi)\:
\left[\hbox{Re}f_{n,n+d}^{(\eta )}(x,\phi)\right]^2       
\:,\label{samp3}
\end{eqnarray}
and likewise for the imaginary part. \\
Quantum tomography opened a fascinating perspective: in fact, there is
the possibility of device-independent measurements of any field-operator,
including the case of generalized observables that do not correspond
to selfadjoint operators as, for example, the complex field amplitude
and the phase. The first application in this direction has been presented in
Ref. \cite{rap} where the number and the phase distributions of a low
excited coherent state have been recovered from the
original tomographic data record. No error estimation was reported in
Ref. \cite{rap}, whereas an earlier analysis of the precision
of such determinations has been reported in Ref. \cite{sen} on the basis
of numerical simulations.
The idea behind these papers is simple. Any field operator $\hat A$, in
fact, is described by its matrix elements $A_{n,m}\equiv \langle 
n|\hat A|m\rangle$ in the number representation. 
Then, upon a suitable truncation of the Hilbert space  dimension, at the
maximum photon number $H$, the expectation value of $\hat A$ is given by
the linear combination
\begin{eqnarray}
\langle \hat A \rangle = \sum_{n,m=0}^{H}  
\overline{\varrho_{n,m}} \ A_{n,m}
\:,\label{samp4}
\end{eqnarray}
whereas the corresponding confidence interval is evaluated by error
propagation calculus
\begin{eqnarray}
\overline{\Delta A}^2 \simeq \sum_{n,m=0}^{H}
\left|\overline{\Delta \varrho_{n,m}}\right|^2 \: \left|A_{n,m}\right|^2
\, .\label{samp5}
\end{eqnarray}
The whole procedure relies on two assumption, namely
\begin{eqnarray}
\varrho_{n,m} \ll 1 \qquad n,m > H
\, ,\label{cond1}
\end{eqnarray}
and      
\begin{eqnarray}
\lim_{n,m\rightarrow\infty} \overline{\Delta\varrho_{n,m}}=0
\, ,\label{cond2}
\end{eqnarray}
which needs a more careful analysis.
The condition in Eq. (\ref{cond1}) is certainly fulfilled for some 
value of $H$, whose determination, however, requires an {\em a priori}
knowledge of the state under examination. On the other hand, it has been
shown in Ref. \cite{err} that in a tomographic measurement involving $N$
experimental data the errors $\hbox{Re}\:\varepsilon_{n,m}$
and $\hbox{Im}\:\varepsilon_{n,m}$ saturate to the value $\sqrt{2/N}$
for $\eta=1$, whereas they diverge exponentially for $\eta < 1$.
Therefore, the condition (\ref{cond2}) cannot be fulfilled in
a real experiment and we conclude that determinations of Ref.
\cite{rap} are not meaningful, as they are affected by diverging errors
and are based on {\em a priori} knowledge of the state. For the same
reason the analysis of Ref. \cite{sen} is not correct, and the added
noise has been largely overestimated.
\section{Homodyning field operators}%
In this paper we analyze the tomographic determination of field-quantities
from a different perspective. By {\em homodyning} an observable $\hat A$ we
mean the average
\begin{eqnarray}
\langle \hat{A} \rangle = \int^{\pi}_0 \frac{d\phi}{\pi} 
\int^{+\infty}_{-\infty} \!\! dx\,p(x;\phi)\:{\cal R}[\hat A](x;\phi) 
\, ,\label{obs1}
\end{eqnarray}
of the state-independent kernel function ${\cal R}[\hat A](x;\phi)$
\cite{tok}, which allows for the determination of the expectation value 
$\langle\hat A\rangle$ without the detour into density matrix elements.
For a Hilbert-Schmidt operator $\hat A$ Eq.(\ref{obs1}) follows
directly from a generalization of Eq.(\ref{bas1}) with 
${\cal R}[\hat A](x;\phi)=\hbox{\rm Tr} \left\{ \hat A K(x-\hat x_{\phi})
\right\}$, whereas alternative approaches to derive explicit expressions
of the kernel have been suggested \cite{ric,tok}, that here we briefly
recall. Starting from the identity involving trilinear products of Hermite
polynomials (valid for $k+m+n=2s$ even \cite{gra} )
\begin{eqnarray}
\int^{+\infty}_{-\infty}\!\! dx e^{-x^2} H_k(x)\,H_m(x)\,H_n(x)= 
\frac{2^{\frac{m+n+k}{2}}\pi^{{1\over2}}k!m!n!}
{(s-k)!(s-m)!(s-n)!}\;,
\end{eqnarray}
Richter proved the following nontrivial formula for the
expectation value of the normally ordered field operators \cite{ric}
\begin{eqnarray}
\langle a^{\dag}{}^n a^m\rangle=\int^{\pi}_0 \frac{d\phi}{\pi}
\int^{+\infty}_{-\infty}\!\! dx\,p(x;\phi)e^{i(m-n)\phi}\frac{H_{n+m}
(\sqrt{2}x)}{\sqrt{2^{n+m}}{{n+m}\choose n}} \nonumber\;,
\end{eqnarray}
which corresponds to the kernel
\begin{eqnarray}
{\cal R}[a^{\dag}{}^n a^m](x;\phi)=e^{i(m-n)\phi}
\frac{H_{n+m}(\sqrt{2}x)}{\sqrt{2^{n+m}}{{n+m}\choose n}}\;.
\end{eqnarray}
For nonunit quantum  efficiency the homodyne photocurrent is rescaled by
$\eta $ whereas the normally ordered expectation  $\langle a^{\dag}{}^n 
a^m\rangle$ gets an extra factor $\eta^{{1\over2} (n+m)}$.  
Therefore, one has
\begin{eqnarray}
{\cal R}_{\eta}[a^{\dag}{}^n a^m](x;\phi)=e^{i(m-n)\phi}
\frac{H_{n+m}(\sqrt{2\eta}x)}{\sqrt{(2\eta)^{n+m}}{{n+m}\choose n}}\;,
\label{Rnm}
\end{eqnarray}
where the kernel ${\cal R}_{\eta}[\hat O](x;\phi)$ is defined as in
Eq. (\ref{obs1}), but now with the experimental probability distribution
$p_{\eta}(x;\phi)$ for nonunit quantum efficiency $\eta$. From Eq. (\ref{Rnm})
by linearity one can obtain
the kernel ${\cal R}_{\eta}[\hat f](x;\phi)$ for any operator function
$\hat f$ that admits a normal ordered expansion
\begin{eqnarray}
\hat f\equiv f(a,a^{\dag})=\sum_{nm=0}^{\infty}f^{(n)}_{nm}a^{\dag}{}^n a^m\;.
\end{eqnarray}
One obtains
\begin{eqnarray}
{\cal R}_{\eta}[\hat f](x;\phi)&=&\sum_{s=0}^{\infty}
\frac{H_s(\sqrt{2\eta}x)}{s!(2\eta)^{s/2}}\sum_{nm=0}^{\infty}
f^{(n)}_{nm}e^{i(m-n)\phi}n!m!\delta_{n+m,s}\nonumber
\\ &=&
\sum_{s=0}^{\infty}\frac{H_s(\sqrt{2\eta}x)i^s}{s!
(2\eta)^{s/2}}\frac{d^s}{dv^s}\Bigg|_{v=0}\!\!\!\!
{\cal F}[\hat f](v;\phi),\label{123}
\end{eqnarray}
where
\begin{eqnarray}
{\cal F}[\hat f](v;\phi)=\sum_{nm=0}^{\infty}f^{(n)}_{nm}
{{n+m}\choose m}^{-1}(-iv)^{n+m}e^{i(m-n)\phi}\;.\label{ff}
\end{eqnarray}
Continuing from Eq. (\ref{123}) one obtains
\begin{eqnarray}
{\cal R}_{\eta}[\hat f](x;\phi)=\exp\left(\frac{1}{2\eta}\frac{d^2}{dv^2}+
2ix\frac{d}{dv}\right)\Bigg|_{v=0}{\cal F}[\hat f](v;\phi)\;,
\end{eqnarray}
and finally  
\begin{eqnarray}
{\cal R}_{\eta}[\hat f](x;\phi)=\int_{-\infty}^{+\infty}
\frac{dw}{\sqrt{2\pi\eta^{-1}}}e^{-\frac{\eta}{2}w^2}{\cal F}[\hat f]
(w+2ix;\phi)\;.\label{generalf}
\end{eqnarray}
In summary, the operator $\hat f$ possesses a tomographic
kernel ${\cal R}_{\eta}[\hat f](x;\phi)$ if the function ${\cal F}
[\hat f](v;\phi)$ in Eq. (\ref{ff}) grows slower than $\exp (-\eta v^2/2)$ 
for $v\to\infty$. In addition, as we can assume that $p_{\eta}(x;\phi)$ 
goes to zero faster than exponentially at $x\to\infty$, the average in
Eq. (\ref{obs1}) is meaningful for the integral in Eq. (\ref{generalf}) 
growing at most exponentially for $x\to\infty$. 
In the next section we will consider the tomographic determination
of four relevant field quantities: the field intensity, the real field
or quadrature, the complex field, and the phase, for all of which the above
conditions are satisfied.
\section{Added noise in tomographic measurements}\label{s:add}
As already mentioned in the previous section the tomographic measurement
of the quantity $\hat A$ is {\em defined} as the average
$\overline{w_{\eta}}$ of the kernel $w_{\eta}\equiv {\cal R}_{\eta}[\hat A]
(x,\phi )$ over the homodyne data. A convenient measure for the precision of
the measurement is given by the confidence interval $\overline{\Delta
w_{\eta}}$ which, being $w_{\eta}$ a real quantity, is given by
$\overline{\Delta w_{\eta}}= \left\{\overline{w^2_{\eta}}
-\overline{w_{\eta}}^2\right\}^{1/2}$, where
\begin{eqnarray}
\overline{w^2_{\eta}} \equiv \overline{{\cal R}_{\eta}^2[\hat A](x,\phi)}=
\int^{\pi}_{0} \frac{d\phi}{\pi}\int_{-\infty}^{\infty}
\!\! dx\:p_{\eta} (x,\phi)\: {\cal R}^2_{\eta}[\hat A](x,\phi)\:.
\label{ysquared}
\end{eqnarray}
When the quantity $\hat A$ can also be directly measured by a specific setup
it makes sense to compare tomographic precision $\overline{\Delta w}$ with
the corresponding fluctuations $\sqrt{\langle\widehat{\Delta A^2}\rangle}$. 
Notice that, when we deal with $\eta <1 $ the noise $\sqrt{\langle\widehat{
\Delta A^2} \rangle_{\eta}}$ is larger that the quantum
fluctuations due to smearing effect of nonunit quantum efficiency.
As we will see, the  tomographic measurement is always more noisy than the
corresponding direct measurement for any observable, and any quantum
efficiency $\eta$. However, this is not surprising, in view of the larger amount
of information retrieved in the tomographic measurement compared to the
direct measurement of a single quantity. \\
In Table \ref{t:one} we report the tomographic quantities $w_{\eta}$
the field-observables examined. Before going into details of each
observable, we mention a useful formula for evaluating confidence intervals.
These are obtained by averaging quantities like
\begin{eqnarray}
{\cal R}_{\eta}^2 [a^{\dag n} a^m](x,\phi)= \label{er1} 
e^{2i\phi(m-n)} \frac{H^2_{n+m}\left(\sqrt{2\eta}x\right)}
{(2\eta)^{(n+m)}
\left(\begin{array}{c}m+n \\ m\end{array}\right)^2}
\:.
\end{eqnarray}
By means of the following identity for the Hermite polynomials \cite{wun}
\begin{eqnarray}
H^2_{n}(x)= 2^n n!^2 \sum_{k=0}^{n}\frac{H_{2k}(x)}{k!^2\:2^k\:(n-k)!}
\label{er2}\:,
\end{eqnarray}
we arrive at
\begin{eqnarray}
\hspace{-15pt}
{\cal R}^2_{\eta} [a^{\dag n} a^m](x,\phi) = e^{2i\phi(m-n)}
\frac{n!^2 m!^2}{\eta^{m+n}} 
\sum_{k=0}^{m+n}\frac{(2k)!\eta^k}{k!^4
(n+m-k)!}{\cal R}_{\eta}[a^{\dag k} a^k](x,\phi)
\label{er3}\:, 
\end{eqnarray}
which expresses the generic square kernel 
${\cal R}^2_{\eta} [a^{\dag n} a^m](x,\phi)$ in terms of "diagonal" kernels 
${\cal R}_{\eta}[a^{\dag k} a^k](x,\phi)$ only.
\subsection{Field-Intensity}
Photodetection is the direct measurement of the field-intensity.
For a single-mode of the radiation field it corresponds to the number
operator $\hat n= a^{\dag} a$.
For nonunit quantum efficiency $\eta$ at the photodetectors,
only a fraction of the incoming photons
is revealed, and the probability of detecting $m$ photons is given by
the Bernoulli convolution
\begin{eqnarray}
p_{\eta}(m) = \sum_{n=m}^{\infty} \rho_{nn}
\left(\begin{array}{c} n \\ m \end{array}\right)
\eta^m (1-\eta)^{n-m}
\label{pho1}\:,
\end{eqnarray}
$\rho_{nn}$ being the actual photon number distribution of the mode
under examination. One considers the reduced photocurrent
\begin{eqnarray}
\hat I_{\eta} = \frac{1}{\eta} a^{\dag}a
\label{pho2}\:,
\end{eqnarray}
which is the quantity that traces the photon number, namely it has the
same mean value
\begin{eqnarray}
\langle \hat I_{\eta} \rangle = \frac{1}{\eta}
\sum_{m=0}^{\infty} m\:p(m) = \bar{n}
\label{pho3}\:,
\end{eqnarray}
where we introduced the shorthand notation $\bar{n}=\langle
a^{\dag}a \rangle$. On the other hand, one has
\begin{eqnarray}
\langle\widehat{\Delta I^2}\rangle_{\eta}=\frac{1}{\eta^2}
\sum_{m=0}^{\infty} m^2 p(m) = \langle\widehat{\Delta n^2}\rangle  +
\bar{n} \left(\frac{1}{\eta}-1\right)
\label{pho4}\:.
\end{eqnarray}
In Eq. (\ref{pho4}) $\langle\widehat{\Delta n^2}\rangle$ denotes the 
intrinsic photon number variance . The term $\bar{n} (\eta^{-1} -1)$
represents the noise introduced by inefficient detection.
The tomo\-graphic kernel that traces the pho\-ton number is given
by the pha\-se-inde\-pendent function $w_{\eta}\equiv 2x^2 - (2\eta)^{-1}$.
With the help of Eq. (\ref{er3}) we can easily evaluate its
variance, namely
\begin{eqnarray}
\overline{\Delta w_{\eta}}^2 = \langle\widehat{\Delta n^2}\rangle
+ \frac{1}{2}\langle\widehat{n^2}\rangle + \bar{n} \left(\frac{2}{\eta}
-\frac{3}{2}\right) + \frac{1}{2\eta^2} 
\label{pho5}\:.
\end{eqnarray}
The difference between $\overline{\Delta w_{\eta}}^2$ and
$\langle\widehat{\Delta I^2}\rangle_{\eta}$ defines the noise 
$N[\hat n]$ added by tomographic method in the determination of the field 
intensity
\begin{eqnarray}
N[\hat n] = \frac{1}{2}\left[ \langle\widehat{n^2}\rangle+
\bar{n}\left(\frac{2}{\eta}-1 \right) + \frac{1}{\eta^2}\right]
\label{pho6}\:.
\end{eqnarray}
The noise $N[\hat n]$ added by the tomographic measurement is always a
positive quantity and largely depends on state under examinations.
For coherent states we consider the noise-ratio
\begin{eqnarray}
\delta n_{\eta} = \sqrt{\frac{\overline{\Delta w_{\eta}}^2}{\langle\widehat{
\Delta I^2}\rangle_{\eta}}}=\left\{2+\frac{1}{2}\left(\eta\bar{n}+
\frac{1}{\eta\bar{n}}\right)\right\}^{1/2}        
\label{pho7}\;,
\end{eqnarray}                                   
which is minimum for $\bar{n}=\eta^{-1}$.
\subsection{Real Field}
For a single mode light-beam the electric field is proportional to a
field quadrature $\hat x = 1/2 (a^{\dag}+a)$, which is just traced
by homodyne detection at fixed zero-phase with respect to the local oscillator.
The tomographic kernel, that traces the mean value
$\hbox{Tr}\{\hat\varrho\hat x\}$ is given by $w_{\eta}\equiv {\cal
R}_{\eta} [\hat x ](x,\phi) = 2  x \cos\phi$, independently on $\eta$,
whereas the square kernel $w_{\eta}^2 \equiv {\cal R}^2[\hat x ](x,\phi)
=4 x^2 \cos^2\phi$ can be rewritten as
\begin{eqnarray}
w_{\eta}^2 =\frac{1}{4}\left[
{\cal R}[a^2](x,\phi) + {\cal R}[a^{\dag 2}](x,\phi)\right]
+{\cal R}[a^{\dag} a](x,\phi) + \frac{1}{2\eta}
\label{fie1}\;.
\end{eqnarray}    
The confidence interval is thus given by 
\begin{eqnarray}
\hspace{-20pt}
\overline{\Delta w_{\eta}}^2 =
\frac{1}{4}\left[\langle a^{\dag 2} \rangle
+ \langle a^2\rangle\right] + \bar{n} + \frac{1}{2\eta} - \left\langle 
\frac{a+a^{\dag}}{2}\right\rangle^2 = \langle\widehat{\Delta x^2}
\rangle + \frac{1}{2}\bar{n} +\frac{2-\eta}{4\eta}
\label{fie11}\;, 
\end{eqnarray}
$\langle\widehat{\Delta x^2 }\rangle$ being the intrinsic quadrature
fluctuations. For coherent states Eq. (\ref{fie11}) reduces  to
\begin{eqnarray}
\overline{\Delta y}_{\eta}^2 = \frac{1}{2}\left[ \bar{n}+\frac{1}{\eta}
\right]
\label{fie12}\;, 
\end{eqnarray}
The tomographic noise in Eq. (\ref{fie11}) has to be compared with the
rms variance of homodyne detection for nonunit quantum efficiency.
This is given by
\begin{eqnarray}
\langle\widehat{\Delta x^2 }\rangle_{\eta} =
\langle\widehat{\Delta x^2} \rangle +\frac{1-\eta}{4\eta}
\label{fie20}\;,
\end{eqnarray}    
For coherent states Eq. (\ref{fie20}) becomes  
$\langle\widehat{\Delta x^2 }\rangle_{\eta} = 1/(4\eta) $
The added noise results
\begin{eqnarray}
 N[\hat x]= \frac{1}{2}\left[\bar{n}+\frac{1}{2\eta}\right]
\label{fie2}\;, 
\end{eqnarray}
whereas the noise-ratio for coherent states is given by
\begin{eqnarray}
\delta x_{\eta} = \sqrt{\frac{\overline{\Delta w_{\eta}}^2}{\langle\widehat{
\Delta x^2}\rangle_{\eta}}} =\left\{2\left(1+\eta\bar{n}\right)\right\}^{1/2} 
\label{fie3}\;,
\end{eqnarray}        
and increases with the scaled intensity $\eta\bar{n}$.
\subsection{Field amplitude}
The detection of the complex field amplitude of a single-mode light-beam 
is represented by the generalized measurement of the annihilation operator 
$a$. The tomographic kernel for $a$ is given by the complex function 
$w_{\eta}\equiv{\cal R}[a](x,\phi) = 2x \exp(i\phi)$. To evaluate the
precision of the measurement one has to consider the noise of a
complex random variable. Generally there are two noises
\begin{eqnarray}
\overline{\Delta w_{\eta}}^2 =
\frac{1}{2}\left[\overline{|w|_{\eta}^2} -
|\overline{w_{\eta}}|^2\pm|\overline{\Delta w_\eta^2}|\right]
\label{amp1}\:,
\end{eqnarray}
corresponding to the eigenvalues of the  covariance matrix. 
Using Eq. (\ref{er3}) one has
\begin{eqnarray}
w_{\eta}^2 \equiv {\cal R}^2_{\eta}[a](x,\phi) = e^{i2\phi}
\left[\frac{1}{\eta}+2{\cal R}_{\eta}[a^{\dag}a ](x,\phi) \right] =
\frac{e^{i2\phi}}{\eta} + {\cal R}_{\eta}[a^2 ](x,\phi) 
\label{amp2}\:.
\end{eqnarray}    
and 
\begin{eqnarray}
\left|w_{\eta}\right|^2 \equiv  \left|{\cal R}_{\eta}[a](x,\phi)\right|^2 =
\frac{1}{\eta} \left[1+2\eta {\cal R}_{\eta}[a^{\dag}a ](x,\phi)\right]
\label{amp3}\:, 
\end{eqnarray}
which lead to
\begin{eqnarray}
\overline{\Delta w_{\eta}}^2 = \frac{1}{2}
\left[\frac{1}{\eta}+ 2 \bar{n} - \left|\langle a \rangle \right|^2
\pm\left| \langle a^2 \rangle- \langle a \rangle^2  \right|
\right]
\label{amp4}\:,
\end{eqnarray}
because $\overline{e^{in\phi}}= \delta_{n0}$ for all states.
The optimal measurement of the complex field $a$, corresponding to the
joint measurement of any pair of conjugated quadratues $\hat x_{\phi}$
and $\hat x_{\phi+\pi/2}$ can be accomplished in a number of different
ways: by heterodyne detection \cite{het}, eight-port homodyne detection
\cite{wal,hai,rip}, or by six-port homodyne detection \cite{zuc,tri}.
In such devices each experimental event $\alpha=x+iy$ in the complex plane
consists of a simultaneous detection of the two commuting photocurrents
$\hat x$ and $\hat y$, which in turn trace the pair of
field-quadratures. The probability distribution is represented by the
generalized Wigner function $W_{s} (\alpha,\bar\alpha)$
with ordering parameter $s$ related to the quantum efficiency as
$s=1-2\eta^{-1}$. The precision of such measurement is defined likewise
Eq. (\ref{amp1}) as follows
\begin{eqnarray}
\langle\widehat{\Delta a^2}\rangle _{\eta}= 
\frac{1}{2}\left[\overline{|\alpha|^2} - |\overline{\alpha}|^2
\pm \left|\overline{\alpha^2} -\overline{\alpha}^2\right|\right]     
\label{amp41}\:,
\end{eqnarray}
where 
\begin{eqnarray}
\overline{\alpha}     &=& \int_{\bf C} d^2\alpha \: \alpha           \:
W_{s} (\alpha,\bar\alpha) =\langle a \rangle   \nonumber \\
\overline{\alpha^2}   &=& \int_{\bf C} d^2\alpha \: \alpha^2         \:
W_{s} (\alpha,\bar\alpha) =\langle a^2 \rangle \nonumber \\
\overline{|\alpha|^2} &=& \int_{\bf C} d^2\alpha \: \alpha\alpha^{*} \:
W_{s} (\alpha,\bar\alpha) =\langle a^{\dag} a \rangle + \frac{1}{\eta}
\label{amp5}\:.
\end{eqnarray}         
>From Eqs. (\ref{amp41}) and (\ref{amp5}) we have
\begin{eqnarray}
\langle\widehat{\Delta a^2}\rangle _{\eta}= 
\frac{1}{2}\left[\bar{n}+\frac{1}{\eta} - |\langle a\rangle|^2
\pm\left|\langle a^2\rangle -\langle a\rangle^2 \right|\right]     
\label{amp51}\:,
\end{eqnarray}
The noise added by quantum tomography is thus
simply given by 
\begin{eqnarray}
N[a] = \frac{1}{2}\bar{n} 
\label{amp6}\:,  
\end{eqnarray}           
which is independent on quantum efficiency. \\
For a coherent state we have 
\begin{eqnarray}
\overline{\Delta w_{\eta}}^2 = \frac{1}{2} \left[\bar{n} +
\frac{1}{\eta} \right]  \quad \hbox{\rm and } \quad
\langle\widehat{\Delta a^2}\rangle _{\eta}= \frac{1}{2\eta}
\label{amp7}\:,  
\end{eqnarray}           
and the noise ratio is given by
\begin{eqnarray}
\delta a_{\eta} = \sqrt{\frac{\overline{\Delta w_{\eta}}^2}{\langle\widehat{
\Delta a^2}\rangle_{\eta}}} =\left\{1+\eta\bar{n}\right\}^{1/2}          
\label{amp8}\;.
\end{eqnarray}         
\subsection{Phase}
\begin{figure}[h]
\begin{center}\psfig{file=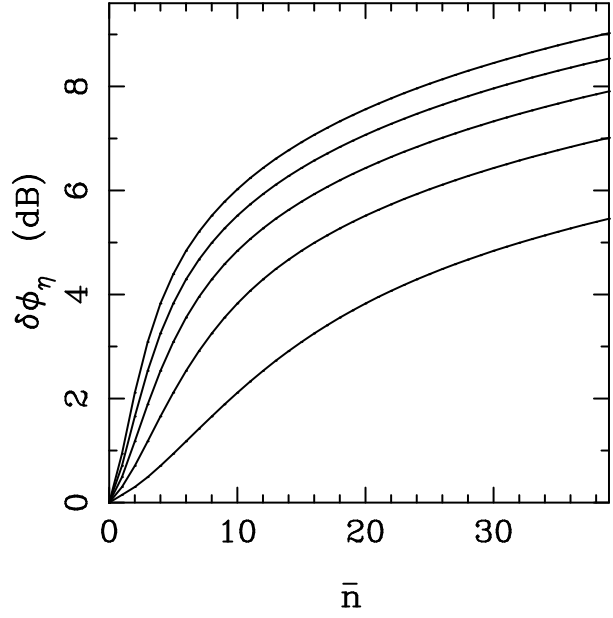,width=8cm}\end{center}       
\caption{Ratio between tomographic and heterodyne noises in the
measurement of the phase for low excited coherent states, The noise
ration is reported versus the mean photon number $\bar{n}$ for some
values of the quantum effiency. From bottom to top we have
$\eta=0.2,0.4,0.6,0.8,1.0$.}\label{f:pha}
\end{figure}

The canonical description of the quantum optical phase is given by the 
probability operator measure \cite{hol,hel}
\begin{eqnarray}
d\mu(\phi) = \frac{d\phi}{2\pi} \sum_{n,m=0}^{\infty}
\exp\{i(m-n)\phi\} |n\rangle\langle m|
\label{pha1}\:,
\end{eqnarray}
which defines a phase operator \cite{pop} through the relation
\begin{eqnarray}
\hat \phi = \int_{-\pi}^{\pi} d\mu(\phi)\: \phi = 
-i \sum_{n \neq m} (-)^{n-m} \frac{1}{n-m} \; \vert n\rangle \langle m\vert 
\label{pha2}\:.
\end{eqnarray}
In principle, a comparison between homodyne tomography and direct
determination of the phase would require from one side the
average of the kernel corresponding to the operator $\hat\phi$, and from
the other side the direct experimental sample of the operator
$\hat\phi$. However, such a comparison would be purely academic, as
there is no feasible setup achieving the optimal measurement (\ref{pha1}).
For this reason, here we consider the heterodyne measurement of the
phase, and compare it with the phase of the tomographic kernel for the
corresponding field operator $a$, i.e. $w_{\eta}=\arg(2xe^{i\phi})$.
Notice that the phase $w_{\eta}$ is not just the given local oscillator
phase, because x has varying sign. Hence averaging $w_{\eta}$ is
not just the trivial average over the scanning phase $\phi$. The
probability distribution of such kernel variable
can be easily obtained by the following identity
\begin{eqnarray}
\int_{0}^{\pi} \frac{d\phi}{\pi}\int_{-\infty}^{\infty} \!\! dx\: 
p_{\eta} (x,\phi)= 1 = \int_{-\pi}^{\pi} 
\frac{dw_{\eta}}{\pi}\int_{0}^{\infty} \!\! dx\: p_{\eta} (x,w_{\eta})
\label{pha4}\:, 
\end{eqnarray}
which implies
\begin{eqnarray}
p_{\eta} (w_{\eta}) = \frac{1}{\pi} \int_{0}^{\infty} \!\! dx \:
p_{\eta} (x,w_{\eta})
\label{pha5}\:. 
\end{eqnarray}
The precision in the tomographic phase measurement is given by the rms
variance $\overline{\Delta w_{\eta}}^2$ of the probability (\ref{pha5}).
In the case of a coherent state $|\beta\rangle \equiv |
|\beta|\rangle$ (zero mean phase) Eq. (\ref{pha5}) becomes
\begin{eqnarray}
p_{\eta} (w_{\eta}) = \frac{1}{2\pi} \left\{1+\hbox{\rm Erf}
\left[\frac{\sqrt{2}|\beta| \cos w_{\eta}}{\sqrt{\eta}}\right]\right\}
\label{pha6}\:, 
\end{eqnarray}
which approaches a "boxed" distribution in $[-\pi/2,\pi/2]$ for 
large intensity.
We compare the tomographic phase measurement with its heterodyne detection,
namely the phase of the detected complex field $a$.
The outcome probability distribution is the marginal distribution
of the generalized Wigner function $W_{s} (\alpha,\bar\alpha)$
($s=1-2\eta^{-1}$) integrated over the radius 
\begin{eqnarray}
p_{\eta}(\phi )= \int_0^{\infty} \! \rho\: d\rho\; W_{s}(\rho e^{i\phi},
\rho e^{-i\phi}) \label{pha3}\:, 
\end{eqnarray}                           
whereas the precision in the phase measurement is given by its rms 
variance $\overline{\Delta \phi}_{\eta}^2$.
We are not able to give a closed formula for the added noise
$N[\phi]=\overline{\Delta w_{\eta}}^2-\overline{\Delta\phi}_{\eta}^2$.
However, for high excited coherent states $|\beta\rangle \equiv |
|\beta|\rangle$ (zero mean phase) one has $\overline{\Delta y}_{\eta}^2
= \pi^2/12$ and $\overline{\Delta \phi}_{\eta}^2= (2\eta\bar{n})^{-1}$.
The asymptotic noise-ratio is thus given by
\begin{eqnarray}
\delta \phi_{\eta}= \sqrt{\frac{\overline{\Delta y}_{\eta}^2}{\overline{
\Delta \phi}_{\eta}^2}} = \pi \sqrt{\frac{\eta\bar{n}}{6}}     
\qquad \bar{n} \gg 1
\label{pha7}\;.
\end{eqnarray}         
A comparison for low excited coherent states can be performed
numerically. The noise ratio $\delta\phi_{\eta}$ (expressed in dB) 
is  shown in Fig. \ref{f:pha} for some values of the quantum
efficiency $\eta$. It is apparent that the tomographic determination of 
the phase is more noisy than the heterodyne one also in this
low-intensity regime.
\section{Summary and Remarks}\label{s:out}
\begin{figure}[h]
\begin{center}\psfig{file=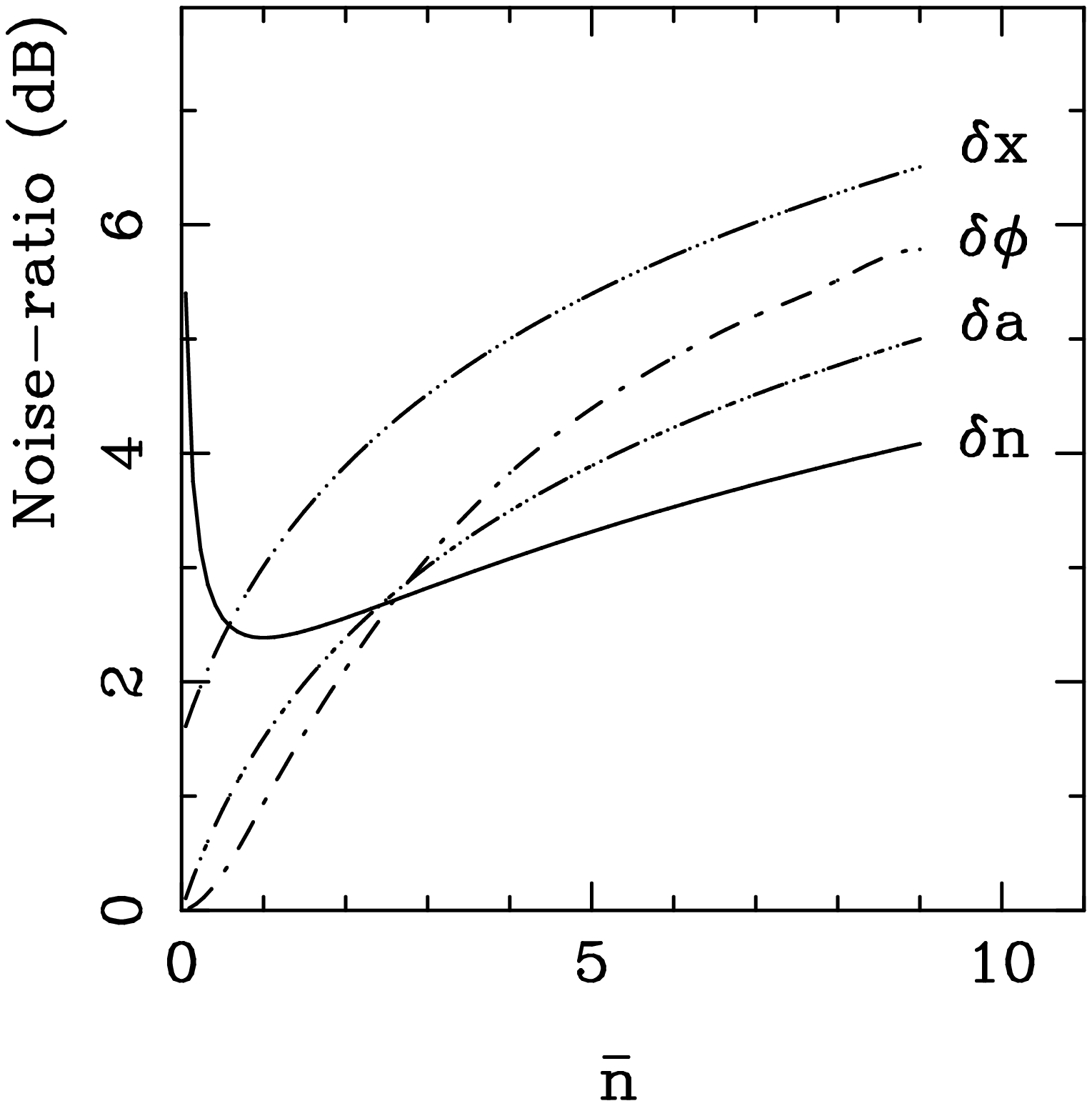,width=8cm}\end{center}       
\caption{The noises ratio (in dB) for all the quantities considered in
this paper. They are plotted for unit quantum efficiency
versus $\bar{n}$: this plot has to compared with Fig. 6
of Ref. \protect\cite{sen}.}\label{f:cmp}
\end{figure}

Homodyne tomography provides a complete characterization of the state 
of the field. By averaging suitable kernel functions it is
possible to recover the mean value of essentially any desidered
field-operator. In this paper we analyzed the determination of the most
relevant observables: intensity, real and complex field, phase.
We have shown that these determinations are affected by noise, which is 
always larger than the corresponding one from  the direct
detection of the considered observables.
In Table \ref{t:two} a synthesis of our results is reported.  \\
We have considered the ratio between the tomo\-graphic and the
di\-rect-measu\-rement noises. This is an increasing function of mean
photon number $\bar{n}$, however scaled by the quantum efficiency
$\eta$. Therefore homodyne tomography turns out to be a very robust
detection scheme for low quantum efficiency.
In Fig. \ref{f:cmp} the noises ratio (in dB) for
all the considered quantities are plotted for unit quantum efficiency
versus $\bar{n}$:  this plot has to compared with Fig. 6
of Ref. \cite{sen}, where the tomographic errors were largely
overestimated. \\
In conclusion, homodyne tomography adds larger noise for highly excited
states, however, it is not too noisy in the quantum regime of low
$\bar{n}$. It is then a matter of convenience to choose between a
direct measurement and homodyne tomography, as the former is the most
precise measurement of the desidered quantity, whereas the latter
represents the best compromise between the conflicting requirements of a
precise and complete measurement of the state of radiation.

\newpage
\begin{table}[h]
\caption{Tomographic vs. direct quantities for the variables of interest
in this paper.}\label{t:one}
\begin{tabular}{lll}
{\sc VARIABLE} & {\sc TOMOGRAPHIC QUANTITY} & {\sc DIRECT QUANTITY} \\
\hline
Intensity&$w_{\eta}\equiv 2x^2-\frac{1}{2\eta}$&$I=a^{\dag} a$\\
Real Field& $w_{\eta}\equiv 2x\cos\phi$&$\hat x= \frac{1}{2}(a+ a^{\dag})$\\
Complex Amplitude&$w_{\eta}\equiv 2x\exp\{i\phi\}$&$a=\hat x+i\hat y $\\
Phase&$w_{\eta}\equiv \arg (xe^{i\phi})$&$\phi=\arg (a)$\\
\hline
\end{tabular}
\end{table}
\vspace{30pt}
\begin{table}[h]
\caption{Added noise in tomographic determinations and noise ratio for
coherent states. For the phase the results are valid in the asymptotic 
regime ${\bar{n} \gg 1}$ }\label{t:two}
\begin{tabular}{lll}
{\sc VARIABLE} & {\sc ADDED NOISE } & {\sc NOISE RATIO} \\
\hline
Intensity&$\quad N[\hat n] = \frac{1}{2}\left[\langle\widehat{n^2}\rangle+
\bar{n}\left(\frac{2}{\eta}-1\right)+\frac{1}{\eta^2}\right]$&$\delta n_{\eta}
=\left\{2+\frac{\eta\bar{n}}{2}+\frac{1}{2\eta\bar{n}}\right\}^{1/2}$ \\
Real Field&$\quad N[\hat x]=\frac{1}{2}\left[\bar{n}+\frac{1}{2\eta}\right]$
&$\delta x_{\eta}=\left\{2\left(1+\eta\bar{n}\right)\right\}^{1/2}$\\
Complex Amplitude&$\quad N[a]=\frac{1}{2}\bar{n}$&$\delta a_{\eta}=\left\{
1+\eta\bar{n}\right\}^{1/2}$\\
Phase &$\quad N[\phi]=\frac{\pi}{12}-
\frac{1}{2\eta\bar{n}}$&$\delta\phi_{\eta}=\pi\sqrt{\frac{\eta\bar{n}}{6}}$\\
\hline\end{tabular}
\end{table}
\newpage

\begin{thebibliography}{99}
\bibitem{ray} D. T. Smithey, M. Beck, M. G. Raymer, A. Faridani, Phys. Rev. Lett. {\bf 70}, 1244 (1993).
\bibitem{tom} G. M. D'Ariano, C. Macchiavello, M. G. A. Paris, Phys. Rev. A{\bf 50} 4298 (1994).    
\bibitem{ulf} G. M. D'Ariano, U. Leonhardt, H. Paul, Phys. Rev. A{\bf 52}, R1801,(1995).        
\bibitem{dar} G. M. D'Ariano, in in {\em Concepts and Advances in Quantum Optics and Spectroscopy of Solids}, ed. by T. Hakioglu  and A. S. Shumovsky. (Kluwer, Amsterdam 1996
\bibitem{rap} D. T. Smithey, M. Beck, J. Cooper, and M. G. Raymer, Phys. Rev. A {\bf 48}, 3159 (1993).         
\bibitem{sen} G. M. D'Ariano, C. Macchiavello, M. G. A. Paris, Phys. Lett. A{\bf 195}, 31 (1994).         
\bibitem{err} G. M. D'Ariano, C. Macchiavello, N. Sterpi, J. Mod. Opt., to appear
\bibitem{tok} G. M. D'Ariano, TOKYO
\bibitem{gra} I. S. Gradshteyn, I. M. Ryzhik, {\em Table of integral, series, and product}, (Academic Press, 1980).       
\bibitem{ric} Th. Richter, Phys. Lett. A {\bf 221} 327 (1996).
\bibitem{wun} A. Orlowsky, A. W\"{u}nsche, Phys. Rev. A{\bf 48} 4617 (1993).
\bibitem{het} J. H. Shapiro, S. S. Wagner, IEEE J. Quantum Electron. QE{\bf20}, 803 (1984); H. P. Yuen, J. H. Shapiro, IEEE Trans. Inform. Theory IT{\bf 26}, 78 (1980).
\bibitem{wal} N.G. Walker, J.E. Carrol, Opt. Quantum Electr. {\bf 18}, 355(1986); N. G. Walker, J. Mod. Opt. {\bf 34}, 15 (1987).
\bibitem{hai} Y. Lay, H. A. Haus, Quantum Opt. {\bf 1}, 99 (1989).
\bibitem{rip} G. M. D'Ariano, M. G. A. Paris, Phys. Rev. {\bf 49} 3022 (1994).    
\bibitem{zuc} A. Zucchetti, W. Vogel, D.-G. Welsch, Phys. Rev. A{\bf 54} 856 (1996)
\bibitem{tri} M. G. A. Paris, A. Chizhov, O. Steuernagel, Opt. Comm. {\bf 134}, 117 (1997).    
\bibitem{hol} A. S. Holevo, {\it Probabilistic and Statistical Aspects of Quantum Theory} (North-Holland Publishing, Amsterdam, 1982).     
\bibitem{hel} C. W. Helstrom, {\em Quantum Detection and Estimation Theory} (Academic Press, New York, 1976).                
\bibitem{pop} Popov V P and Yarunin V S 1992 {\em J. Mod. Opt.} {\bf 39} 1525 
\end{thebibliography}
\end{document}